\documentclass[final]{pasj00}
\SetRunningHead{T. Akahori and K. Masai}{Core Structure of Intracluster Gas}
\title{Core Structure of Intracluster Gas: Effects of Radiative Cooling on Core
Sizes}
\author{Takuya \textsc{Akahori} and Kuniaki \textsc{Masai}}
\affil{ Department of physics, Tokyo Metropolitan University, Hachioji, Tokyo
192-0397}
\email{akataku@phys.metro-u.ac.jp, masai@phys.metro-u.ac.jp}
\Received{2005/8/29}
\Accepted{2006/4/8}
\Published{$\langle$publication date$\rangle$}

\begin{document}

\maketitle

\begin{abstract}

We investigate the core structure of radiatively cooling intracluster gas,
using a hydrodynamics code. We calculate evolution of model clusters of the
initial core radii 160--300~kpc until the initial central cooling time, and
analyze the resultant clusters using the double $\beta$-model as done by
observational studies.  It is found that the core-size distribution thus
obtained shows two peaks $\sim 30$--100~kpc and $\sim 100$--300~kpc and
marginally can reproduce the observed distribution which exhibits two
distinct peaks around $\sim 50$~kpc and $\sim 200$~kpc.  This result may
suggest radiative-cooling origin for small cores, while cooling is yet
insignificant in the clusters of large cores.  It should be noted that the small core peak is reproduced by clusters that are still keeping quasi-hydrostatic balance before the initial central cooling time has elapsed. 

\end{abstract}

\section{Introduction} 

Profiles of X-ray emitting hot gas in clusters of galaxies have been studied often by using the so-called $\beta$-model, which is an isothermal hydrostatic gas model consisting of a core and envelope.  From the $\beta$-model analyses of 121 clusters including nearby clusters (\cite{Mohr99}), Ota and Mitsuda (2002) obtained an interesting result that the distribution of the core radii of the intracluster gas exhibits two distinct peaks at $\sim 50$~kpc and $\sim 200$~kpc for $H_0=70~{\rm km~s^{-1}~Mpc^{-1}}$ (see also \cite{OM04}). 
Akahori and Masai (2005) (hereafter AM05) investigated correlations of the core radii with various properties of the clusters, and found that the radii in the larger core group around the peak $\sim 200$~kpc are marginally proportional to the virial radii of the clusters and therefore the origin may be attributed to simple self-similar collapse.  On the other hand,
in the smaller core group around $\sim 50$~kpc, no clear correlation is found between the core and virial radii, suggesting some other origin of the small core formation.

AM05 investigated several possibilities of the origin of the small cores, not
only by examining correlations among the observational quantities but by
simulating the $\beta$-model with a hydrodynamics code. Many of the small core
clusters possess central cD's or giant ellipticals, but the simulation shows
that the gas core under their presence is $\sim 40$~kpc at most, and is too
small to account for the observed range $\sim 40$--80~kpc.
AM05 suggested another possibility: the effects of radiative cooling on the core
size. Radiative cooling time is shorter in the core than in the ambient region,
because the density is higher.  As the core cools, the ambient gas then could
inflow to compensate the pressure loss inside.  Although the ASCA, Chandra and
XMM-Newton observations (e.g., \cite{Makishima01}; \cite{Lewis02};
\cite{Peter01}) suggested much smaller amount of cooled mass than expected for
the classical cooling flow model (see \cite{Fabian94} for a review), this
process increases the gas density toward the cluster center and is likely
responsible for the small cores observed. AM05 found that the central cooling
time, $t_{\rm cool}$, is significantly shorter than the Hubble time for 48 out
of 50 small core clusters.

Thermal evolution of clusters has been studied by many authors.
Most of the papers addressed processes that would slow or inhibit the onset of 
radiative cooling instability, such as electron conduction and activities of active galactic nuclei.
Until an elapsed time compareble to $t_{\rm cool}$, however, the gas appears to be cooling rather slowly with the temperature $\sim$ keV at the cluster center.
For example, Ruszkowski and Begelman (2002) studied the heating, conduction, and minimum
temperature in cooling flows with a simple spherical model and showed that line
and free-free cooling in the cluster center leads to slow cooling until the initial central cooling time has elapsed.  At this stage feedback is not yet very important.  
Masai and Kitayama (2004) showed that when the intracluster gas is cooling with 
quasi-hydrostatic balancing, the temperature of the cooling gas appears to approach a 
constant value toward the cluster center.

In the present paper, we investigate thermal evolution of clusters, paying attention
to the core radius of the intracluster gas undergoing radiative cooling.  Our
interest is the origin of the small cores or the observed core-size
distribution which exhibits two distinct peaks.  We simulate cooling
clusters using a hydrodynamics code (AM05), and analyze the core radii
applying the $\beta$-model, on which the observed results are based.  In
section 2 we describe the model.  In section 3, we present the results of
calculations and discuss the properties of the cooling cores.  We give some
concluding remarks in section 4. 

\section{Model and Calculations}

We calculate the evolution of radiatively
cooling intracluster gas which is initially in hydrostatic balance under the
presence of dark matter.  We analyze the resultant clusters using the
$\beta$-model in the same manner as the observations (Ota, Mitsuda 2002):
the core-radius distribution having two distinct peaks was obtained by analyses
using the $\beta$-model.  Since the $\beta$-model is based originally on the
King model of the collisionless matter (see e.g., \cite{Sarazin86}), we apply
primarily the King model for the dark matter and galaxies of a cluster.  We
also mention the results in the case of the NFW dark matter model
(Navarro et al. 1996), because such a cuspy profile has been suggested for dark halo
mergers or cluster formation by numerical simulations.  As for the thermal
evolution of the intracluster gas, the two dark matter models show the similar
results to each other.  In the following we describe our models and
calculations along the King dark matter model.
 
With the gravitational potential $\Phi$ of the total mass within radius $r$,
gravitational balance of a spherically symmetric cluster is written as, 
\begin{eqnarray}\label{eq:1}
\begin{array}{lll}
\Phi&= \displaystyle \frac{kT}{\mu m}
\left(\frac{d\ln \rho_{\rm g}}{d\ln r}+\frac{d\ln T}{d\ln r}\right)& \mbox{:
gas}\\
\\ &= \displaystyle \sigma_{*r}^2
\left(\frac{d\ln \rho_*}{d\ln r}+\frac{d\ln \sigma_{*r}^2}{d\ln r}\right)&
\mbox{: dust}
\end{array}
\end{eqnarray}
where ``dust" means collisionless components, i.e. dark matter
and galaxies, $k$ is the Boltzmann constant, $m$ the proton mass, $\mu$ the
mean molecular weight for which we take $\mu=0.6$, $\rho$ the mass density for
each component, $T$ the gas temperature, and $\sigma_r$ is the radial velocity
dispersion which is equal to the line-of-sight velocity dispersion for the
isotropic dust. If isothermal distribution is attained for every component,
i.e. $d\ln T/d\ln r\simeq 0$ and $d\ln \sigma_{*r}^2/d\ln r\simeq 0$,
\begin{equation}\label{eq:2}
\beta_{\rm prof}
\equiv \frac{d\ln \rho_{\rm g}/d\ln r}{d\ln \rho_*/d\ln r}
\simeq\frac{\sigma_{*r}^2}{kT/\mu m}
\equiv \beta_{\rm spec}
\end{equation}
follows equation (\ref{eq:1}). 

For comparison with the observations of core sizes, in modelling clusters we
apply the density profile function based on the $\beta$-model (\cite{CF76}),
\begin{equation}\label{eq:3}
\rho(r)=\rho_0[1+(r/r_{\rm c})^2]^{-3\beta/2},
\end{equation} where $\rho_0$ is the central density, $r_{\rm c}$ is the core
radius, and $\beta$ is a parameter to represent the envelope slope. For the
consistency with the $\beta$-model for the gas, we adopt the King model for the
collisionless components.  Although the original King model (\cite{King66}) is
not of simple analytic form, the profile may be approximated with
equation~(\ref{eq:3}), which is sometimes called approximate King model (see
e.g., \cite{Sarazin86}). Table~\ref{ta:1} shows the best-fit parameters for the
King profile with equation~(\ref{eq:3}) in the range $r/r_{\rm K}=0.0$--$10.0$,
where $r_{\rm K}$ is the core radius of the King model. 

%%% table 1 here %%%

\begin{table}
\caption{Best-fit parameters for the King profile.}
\label{ta:1}
\begin{center}
\begin{tabular}{cccccc}
\hline\hline
\noalign{\smallskip} Range && \multicolumn{3}{c}{Fitting parameters} \\
\noalign{\smallskip}
\cline{3-5}
\noalign{\smallskip}
$r$/$r_{\rm K}$& & $r_{\rm c*}/r_{\rm K}$ & $\rho_{0*}/\rho_{\rm K}(0)$ &
$\beta_{*}$\\
\noalign{\smallskip}
\hline
\noalign{\smallskip} 0.0--1.0 &&  1.04 & 1.00 & 1.13\\ 0.0--2.0&& 1.05 & 1.00 &
1.14\\ 0.0--5.0&& 1.10 & 1.00 & 1.21\\ 0.0--10.0&& 1.34 & 0.98 & 1.46\\
\noalign{\smallskip}
\hline
\end{tabular}
\end{center}
\end{table}

\begin{figure}
\begin{center}
\FigureFile(77mm,54mm){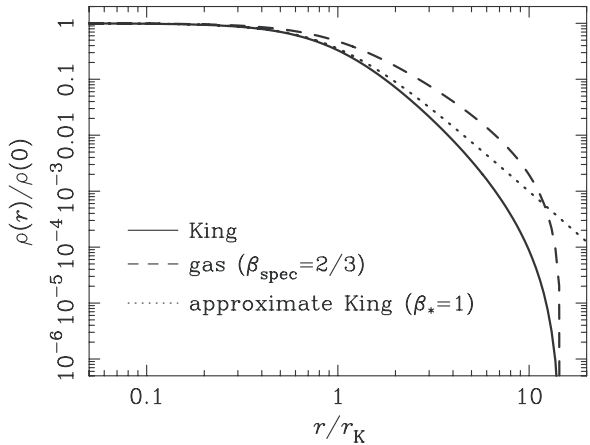}
\end{center}
\caption{Radial density profiles. The solid and dashed lines represent
$\rho_*\propto \rho_{\rm K}$ and $\rho_{\rm g}\propto\rho_{\rm K}^{2/3}$,
respectively. The dotted line is the approximate King profile
(equation~(\ref{eq:3}) with $\beta=1$).  }
\label{fig:1}
\end{figure}

We investigate four models of clusters which consist of dark matter, galaxies,
and gas with their mass ratio $M_{\rm gal}:M_{\rm gas}:M_{\rm DM}=1:5:30$. The
galaxies and dark matter are represented by their gravitational potential
components in the momentum equation (equation~(16) in AM05). We employ the
original King model for their density distributions, and then $\rho_{\rm
g}\propto {\rho_*}^{\beta_{\rm spec}}$ with $\beta_{\rm spec}(\simeq\beta_{\rm
prof})=2/3$ for the initial gas profile (equation~(\ref{eq:2})). The initial
profiles are shown in figure~\ref{fig:1}, where the tidal radius $r_{\rm
t}=r_{\rm vir}$, i.e. $\rho_*(r\ge r_{\rm vir})=0$, and the core radius $r_{\rm
K}=r_{\rm c}$.
The core radius of the gas well reflects the gravitational potential consistently
in the context of the $\beta$-model (AM05).  

%%% figure 1 here %%%

The four model-clusters have different initial core radii of the gas, $r_{\rm
c}=$ 160, 200, 250, and 300~kpc, and satisfy the self-similar relation with their
virial radii, as $x_{\rm vir}\equiv r_{\rm vir}/r_{\rm c}=15$,
or $r_{\rm vir}=$ 2.40, 3.00, 3.75, and 4.50~Mpc (listed in table~\ref{ta:2}). 
These values are on the dashed line in figure 5 of AM05, which represents the
$r_{\rm vir}/r_{\rm c}$ relation obtained for the large core group peaked at
$\sim 200$~kpc in the observed core-size distribution. The initial gas
temperature (isothermal), $T$, under the hydrostatic equilibrium is higher than
the virial temperature, $T_{\rm vir}$, as $\beta_{\rm spec} T\simeq T_{\rm
vir}$ for $x_{\rm vir}^2/(1+x_{\rm vir}^2)\simeq 0.996$ (see AM05).

%%% table 2 here %%%

\begin{table*}
\caption{Initial properties of the simulated clusters.}
\label{ta:2}
\begin{center}
\begin{tabular}{cccccccccc}
\hline\hline
\noalign{\smallskip} & & & &  & & \multicolumn{3}{c}{$\beta$-model
parameters$^{**}$} & \\
\cline{7-9}
\noalign{\smallskip}
\noalign{\smallskip} & & $r_{\rm vir}$ & ${M_{\rm vir}}^{*}$ & $T$ & & $r_c$ &
$n_{\rm g0}$ & $\beta$ & ${t_{\rm cool}}^{**}$\\ Model && (Mpc) &
$(10^{15}\MO)$ & (keV) & & (kpc) & $({\rm 0.01 cm}^{-3})$ & & (Gyr) \\
\noalign{\smallskip}
\hline
\noalign{\smallskip} (a)........ && 2.40 & 0.555 & 3.12 && 171 & 2.15 & 0.83 &
0.689\\ (b)........ && 3.00 & 1.08 & 4.87 && 209 & 2.16 & 0.82 & 0.800\\
(c)........ && 3.75 & 2.62 & 7.61 && 260 & 2.10 & 0.81 & 1.086\\ (d)........ &&
4.50 & 3.66 & 11.0 && 310 & 2.04 & 0.80 & 1.303\\
\noalign{\smallskip}
\hline
\noalign{\smallskip}
\multicolumn{10}{l}{{\small $^{*}$ The virial mass.}}\\
\noalign{\smallskip}
\multicolumn{10}{l}{{\small $^{**}$ Values are estimated after 1.6~Gyr of the
hydrostatic balance test without cooling.}}
\end{tabular}
\end{center}
\end{table*}

We carry out hydrodynamical simulations using a Smoothed Particle Hydrodynamics
(SPH) code (\cite{HK89}; \cite{Monaghan92}) with 95,000 bodies. The smoothing
length is estimated to be $h=17.1$~kpc and $9.97$~kpc for the core region
($n_{\rm g}\sim 0.02~{\rm cm^{-3}}$) and cooled cluster center ($n_{\rm g}\sim
0.1~{\rm cm^{-3}}$), respectively, and sufficiently small compared with the
core size concerned here. Here, $n_{\rm g}=\rho_{\rm g}/\mu m$ is the number
density of the gas. For the general artificial viscosity we adopt the
Monaghan-Gingold value (\cite{MG83}; see also \cite{HK89}) with the
coefficients ($\alpha$, $\beta$) = (1.0, 2.0), and for the gravitational
softening we take $\nabla\Phi\propto 1/(r^2+\epsilon^2)$ with $\epsilon=0.1h$. 

Radiative cooling is taken into account for the energy equation as
\begin{equation}\label{eq:5}
\frac{du_i}{dt}=-\frac{P_i}{\rho _i^2}\nabla\cdot v_i -\rho _i\Lambda(T_i),
\end{equation}
where $u_i$ is the internal energy per unit mass, $v_i$ is the
velocity, and $\Lambda (T_i)$ is the cooling function for the $i$-th
SPH-particle.  No additional heating processes are considered because the
evolution of cooling cores is of our interest.  Neither is electron conduction,
which is not effective within cooling cores.  We use the cooling function of
Sutherland and Dopita (1993), which includes line emission with the metalicity
$Z\sim 0.3 Z_\odot$, by approximation as $\Lambda (T_i)=3.0\times
10^{-27}{T_i}^{1/2}$~${\rm erg~cm^{-3}~s^{-1}}$ in the temperature range
$\sim$1.5--10~keV. 
Actually, until $t\sim t_{\rm cool}$ of practical interest, the temperatures of the
cooling gas are still above $\sim 1.5$~keV where bremsstrahlung dominates.
Here, $t_{\rm cool}=3n_{\rm g0}kT/({n_{\rm g0}}^2\Lambda)$ is the initial
cooling time at the cluster center. 

\begin{figure}
\begin{center}
\FigureFile(77mm,54mm){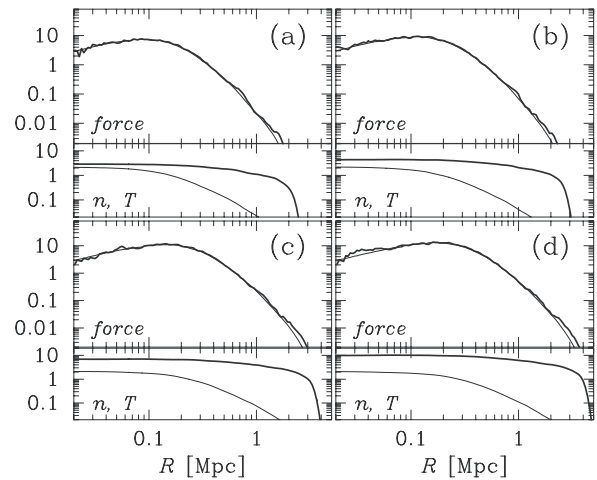}
\end{center}
\caption{ Initial hydrostatic balance of the simulated clusters. We obtain the
clusters after 1.6~Gyr of the hydrostatic balance test without cooling. For
each box, the upper panel represents the gravity on the gas and the pressure
gradient (the thin and thick line, respectively, in a unit $1.79\times
10^{38}~{\rm g~cm~s^{-2}}$), and the lower panel represents the density (the
thin line, $0.01~{\rm cm^{-3}}$) and temperature (the thick line, keV). }
\label{fig:2}
\end{figure}

Before going to calculations of evolution
with radiative cooling, we examine the hydrostatic balance of the gas with
$\Lambda(T_i)=0$ in equation~(\ref{eq:5}), as shown in figure~\ref{fig:2}. 
Gravity on the gas and the pressure gradient balance with each other, and the
initial gas profile is kept at least until 3.2~Gyr, which is enough long
compared with the dynamical (free-fall) timescale, $t_{\rm d}=(32/3\pi
G\rho)^{1/2}\sim 1.6$~Gyr at the cluster center ($n_{g0}=0.02$~${\rm
cm^{-3}}$).  The gas is kept nearly isothermal so that the profile is well
represented by the $\beta$-model, although the temperature declines somewhat at
$r > 5r_{\rm c}$ due likely to adiabatic expansion of the outermost envelope.
At $r\le 5r_{\rm c}$ of practical interest, the hydrostatic gas is represented
well by the $\beta$-model as follows.  From equations (\ref{eq:2}) and
(\ref{eq:3}), the $\beta$-model should give a relation
\begin{equation}\label{eq:4}
\beta_{\rm prof}=\beta/\beta_{*}\simeq\beta_{\rm spec},
\end{equation}
or $\beta\simeq\beta_{*}\beta_{\rm spec}$. Actually, $\beta \sim
0.80$--0.83 (table~\ref{ta:2}) is consistent with $\beta_{*}\beta_{\rm spec}
\sim 0.8$ for $\beta_{*}\sim 1.2$ (table~\ref{ta:1}) and $\beta_{\rm
spec}=2/3$. 

%%% figure 2 here %%%

\section{Results and Discussion}

\subsection{Evolutions of the $\beta$-Model Parameters}

We apply the double $\beta$-model as well as the single $\beta$-model to
simulated clusters.  The double $\beta$-model is the superposition of two
single $\beta$-models as
 \begin{equation}\label{eq:6}
\rho(r)=\rho_1[1+(r/r_1)^2]^{-3{\beta_1}/2}+\rho_2[1+(r/r_2)^2]^{-3{\beta_2}/2}.
\end{equation}
While in observations the parameters are inferred from the
surface brightness profile, which depends substantially on the density profile
as $\propto \rho^2T^{1/2}$, in our calculations they are obtained
straightforwardly from the density profile.  We obtain the best-fit parameters
of the ``outer" components ($\rho_1$, $r_1$, and $\beta_1$) using the data in
the range $1.0\le r/r_{\rm K}\le 5.0$, and then obtain the ``inner" ones
($\rho_2$ and $r_2$), assuming $\beta_2=\beta_1$ as done by Ota and Mitsuda
(2002; 2004), using the data of $0.0\le r/r_{\rm K}\le 5.0$ including the outer
component with $\rho_1$, $r_1$ and $\beta_1$ fixed.

\begin{figure}
\begin{center}
\FigureFile(68mm,100mm){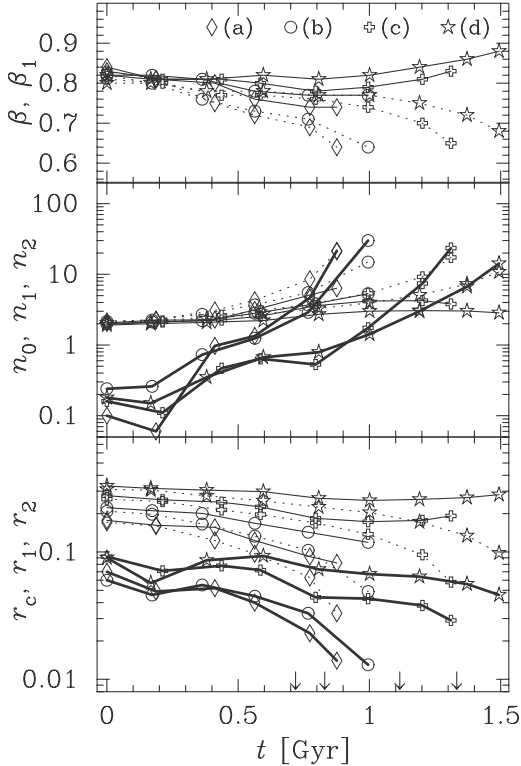}
\end{center}
\caption{Thermal evolutions of the $\beta$-model parameters: slope parameter
(the top panel), number density of center (the middle, in a unit 0.01~${\rm
cm^{-3}}$), and core radius (the bottom, Mpc). The thin and thick solid lines
represent the outer and inner components of the double $\beta$-model,
respectively. The dotted lines represent the components of the single
$\beta$-model. The arrows represent the cooling time of each cluster. }
\label{fig:3}
\end{figure}

Figure~\ref{fig:3} shows the time evolution of the parameters
for the single and double $\beta$-models after the gas starts to cool. We see
in the bottom panel that the outer core radii (the thin solid lines) decrease
very slowly compared with the core radii of the single $\beta$-model (the
dotted lines). This implies that the outer component keeps the initial value
since thermal evolution of the gas is predominant in the inner component.
Actually, $\beta_1$, which is determined by the outer component, is roughly
kept the initial value, although $\beta$ of the single $\beta$-model decreases
by about 10\% within $t_{\rm cool}$ (the top panel).  The behavior is also seen
in the central density.  While the increase in the central density of the outer
component is as small as a factor of $\sim 2$, the central density with the
single $\beta$-model increases exponentially (the middle panel). 

%%% figure 3 here %%%

\begin{figure}
\begin{center}
\FigureFile(77mm,54mm){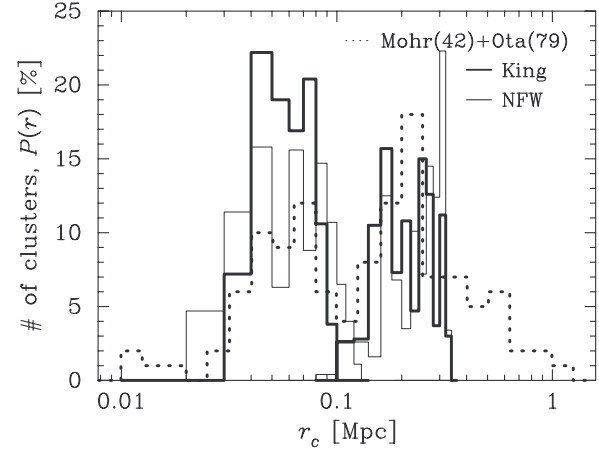}
\end{center}
\caption{Distribution of core radii. The thick solid lines represent the relative
probabilities within $t_{\rm cool}$ of the four simulated clusters, estimated
for the outer and inner cores with the bin of $\Delta r=20$~kpc and 10~kpc,
respectively. The dotted line represents the distribution of core radii of 121
clusters (same as figure~4 of AM05). The thin solid lines represent the relative probabilities for the case of the NFW dark matter potential.}
\label{fig:4}
\end{figure}

From the time evolution of the core radii, we calculate the time during
which a cluster would have the core radius between $r$ and $r+\Delta r$
and estimate the population or relative number of clusters that would fall into
a certain range of core radius.  The core size distribution thus obtained for
the outer and inner cores is shown in figure~\ref{fig:4}. Our calculations
reproduce the observed core-size distribution or two distinct peaks except
for the details such as their widths or the tails. This implies that radiative
cooling is a possible cause of small cores of clusters, while cooling is yet
minor in large core clusters.  It is interesting that even the small core peak
is produced by the clusters of $t \lesssim t_{\rm cool}$, i.e. moderately cooled
clusters that are keeping quasi-hydrostatic balance.

%%% figure 4 here %%%

In the observed distribution of figure~\ref{fig:4} one may notice four clusters
which have very small cores.  It is unlikely that such small cores as
$r_{\rm c}\lesssim 30$~kpc are reproduced by radiative cooling within
$t\lesssim t_{\rm cool}$. Our cooling function underestimates by $\sim 40$~\% at
$T = 1.5$~keV compared with the function of Sutherland and Dopita (1993).
If applied their original function, the small core peak might have a tail toward
the small core end. However, this is unlikely because such a very small core
is quite transient (figure~\ref{fig:3}). This may be related with the fact that clusters
of $t \gtrsim t_{\rm cool}$ are out of quasi-hydrostatic balance.
Another possible explanation of the very small cores is the presence of
central gravitational source; the four clusters evidently have central cD's or giant ellipticals.
The presence of such galaxies is likely responsible for the very small cores (see AM05).

The lack of clusters of core radii $\sim 120$~kpc is seen
in the calculated distribution as well.  Although the outer cores of
$\sim$~100--300~kpc of simulated clusters can explain the population around the
peak of the observed large core clusters, the latter exhibits larger core tail
extended to $r_{\rm c}\sim$~1~Mpc. If we simulate clusters which have
initial huge cores $\gtrsim 0.4$~Mpc, we might get the cooled cores of $\sim
120$~kpc following the self-similarity (see below). However, if such 
huge core clusters are in hydrostatic equilibria as represented by the
$\beta$-model, the clusters would have unlikely virial radii as large as
$r_{\rm vir}\gtrsim 6$~Mpc.  In fact, most of the huge core clusters deviate
from the self-similar relation, as shown in figure~5 of AM05.  

A considerable factor for the origin of such huge cores is mergers.
When merger takes place,  the gas could form a large flat core with a steep envelope
(large $\beta$). Actually, in AM05 sample, 6 out of 7 clusters of $\beta>1$ have cores of
$r_{\rm c}>0.6$~Mpc, and the average of $\beta$ in the 17 clusters (including
14 irregulars) of $r_{\rm c}>0.4$~Mpc is 0.98, which is significantly greater
than the average 0.65 in 121 clusters. 

We find in figure~\ref{fig:3} that the single
$\beta$-model represents the intermediate profile between the inner and outer
profiles of the double $\beta$-model. If cooling is minor and the
self-similarity remains, the profile would be represented by the single $\beta$-model
of a large core.  With cooling the core, but yet minor in the outer, the profile
is being that represented better but transiently by the double $\beta$-model,
and eventually comes to that represented by the single $\beta$-model of a small
core.  As a result, the core-size distribution exhibits two distinct peaks along
the thermal evolution. Two distinct peaks are not seen for the cluster sample of
redshift between 0.4 and 1.3 (\cite{Ettori04}).
The reason may be that their analysis was done with the single $\beta$-model; otherwise,
the cooling might be yet minor in such very distant clusters.
We confirm that two distinct peaks do not appear if applied the single $\beta$-model
to the simulated clusters. 
The estimation of the inner core radius depends on
$\beta_1$ obtained from the outer slope. If $\beta_1$ is taken to be 10~\% smaller than
that in the present calculations, the resultant clusters might have 10~\% smaller cooling
cores, as expected from the $\beta$--$r_{\rm c}$ correlation
described in figure~\ref{fig:1}. Thus, the small-core distribution merely shifts toward 
the smaller side. 

We also investigate clusters of the NFW dust profile (Navarro et al. 1996),
where the parameters of the NFW profile are determined so that the clusters
have nearly the same initial density and temperature profiles of the gas, and analyze
the core sizes in the same manner as above.
The obtained core-radius distribution is similar except for a few \% tail extended to $\sim
20$~kpc (figure~\ref{fig:4}). 
There is no appreciable difference between the NFW and King dark matter models
in the cooling gas distribution. 

%%% figure 5 here %%%

\begin{figure*}
\begin{center}
\FigureFile(46mm,30mm){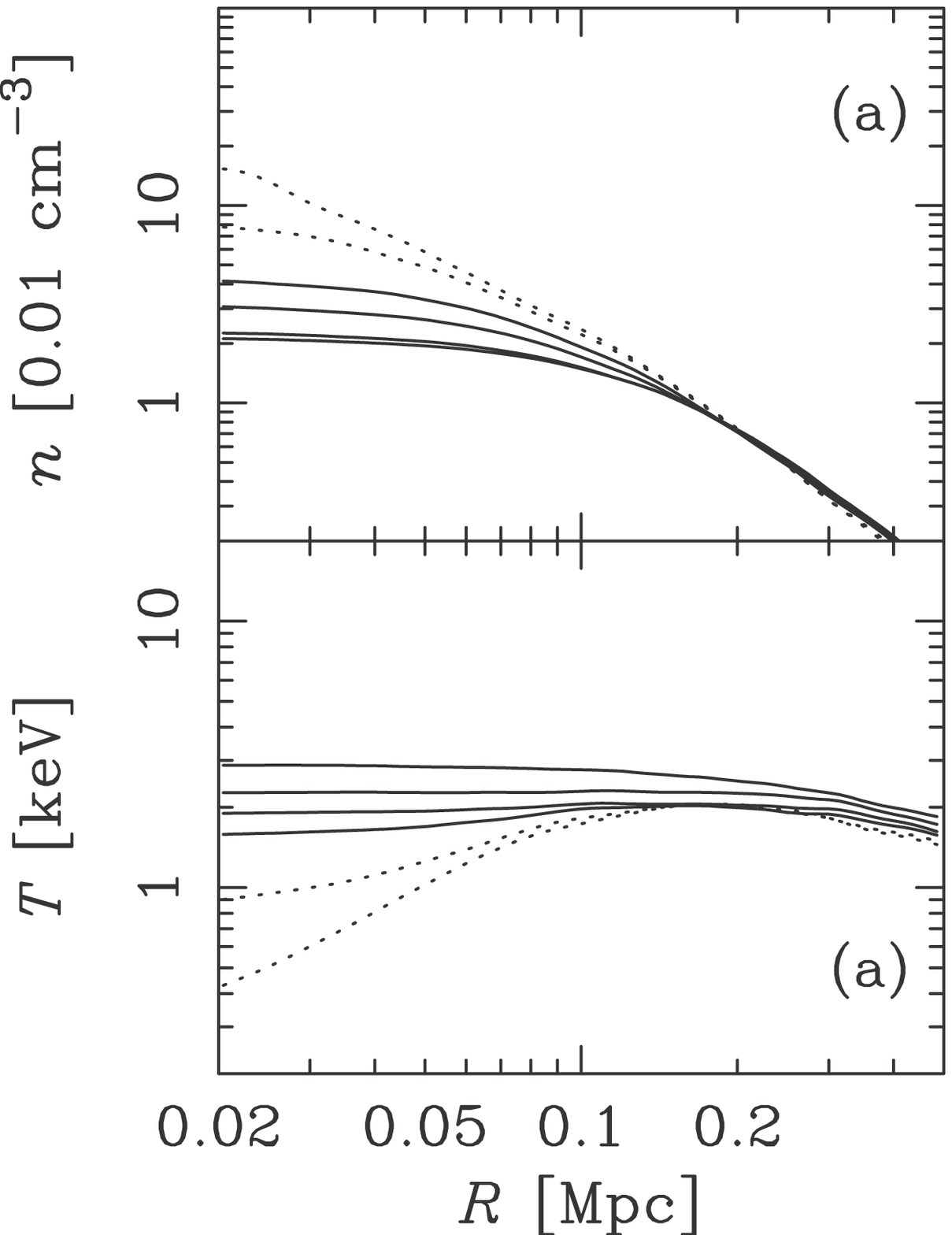}
\FigureFile(38mm,27mm){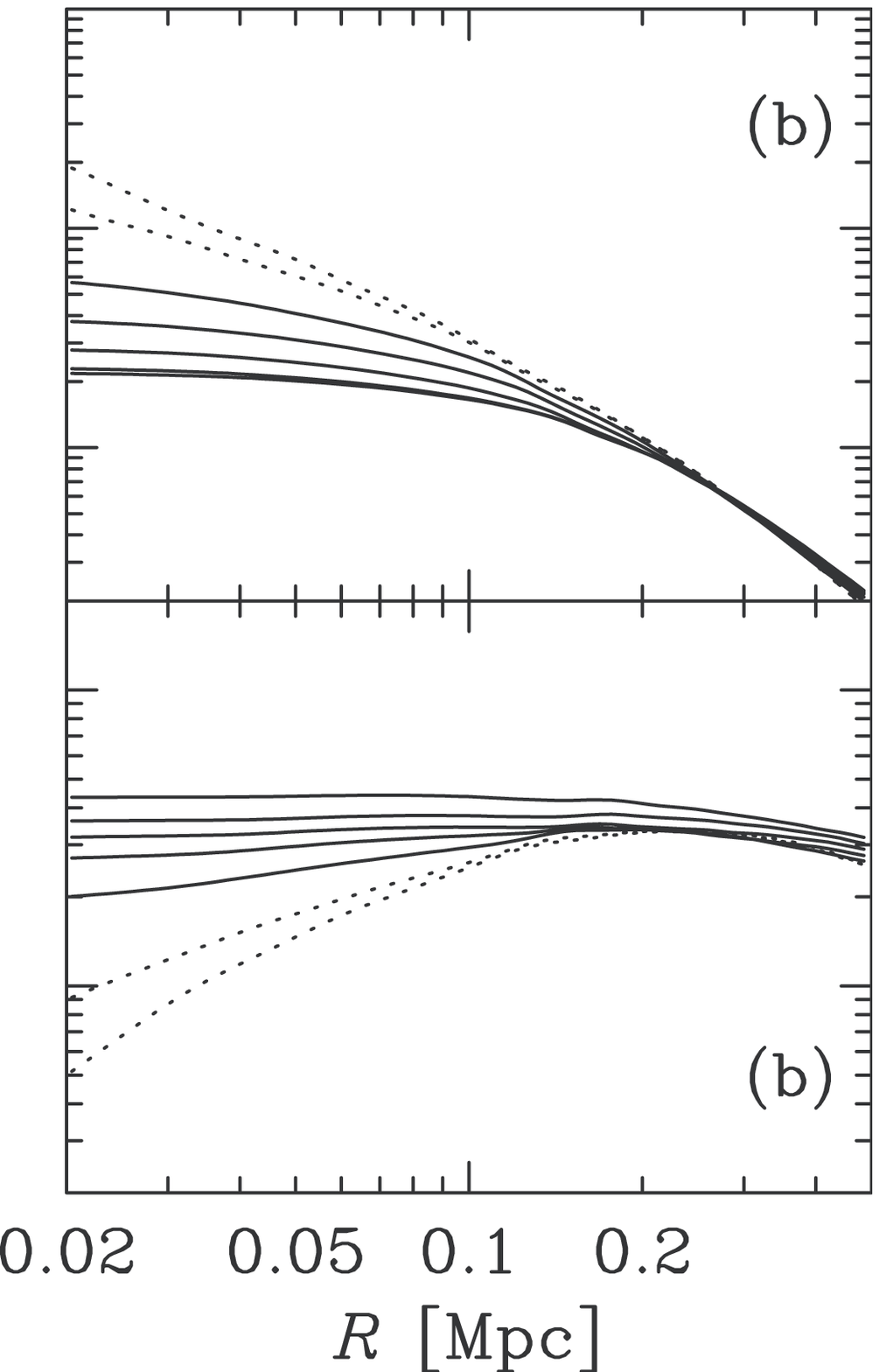}
\FigureFile(38mm,27mm){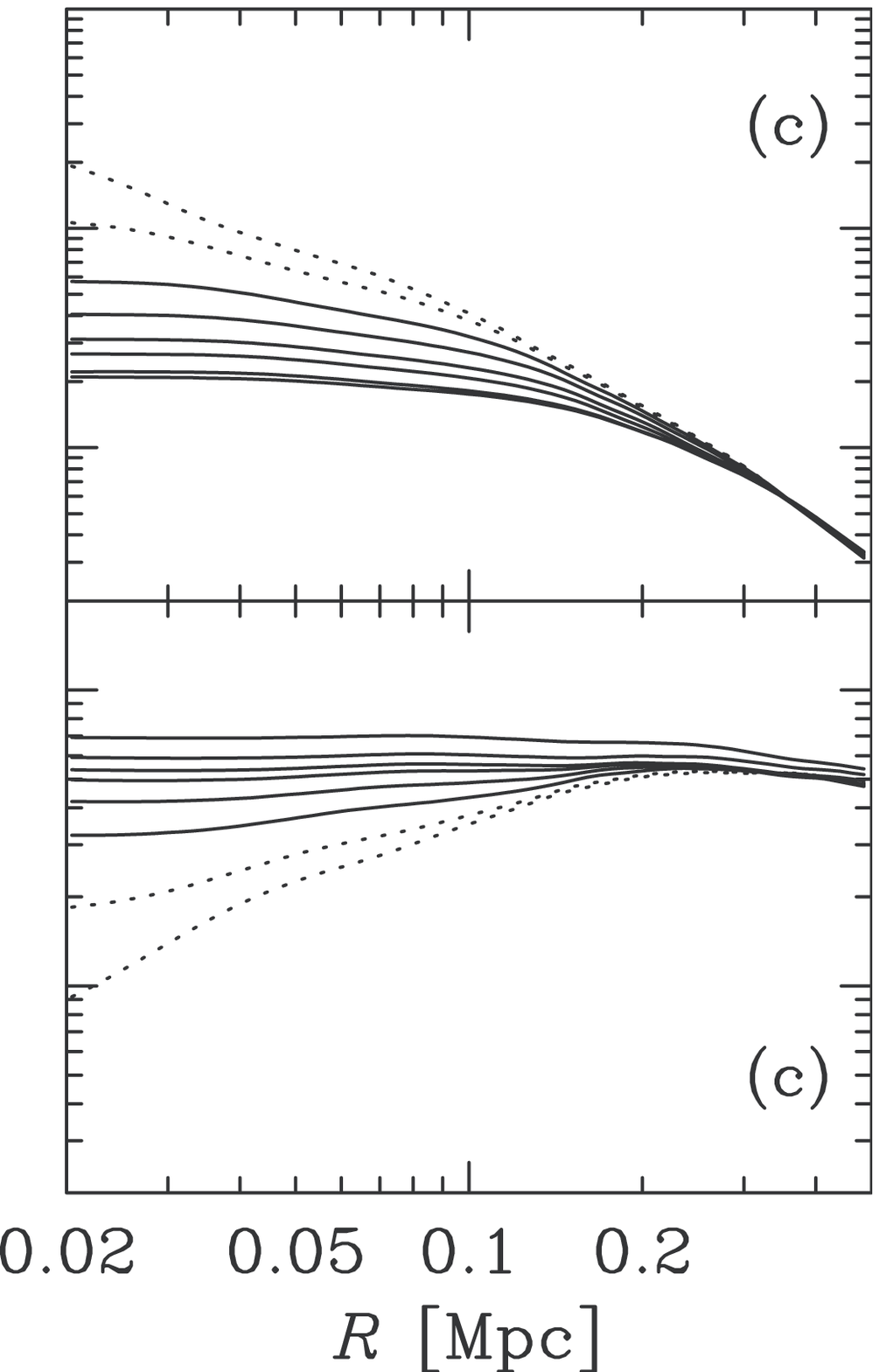}
\FigureFile(38mm,27mm){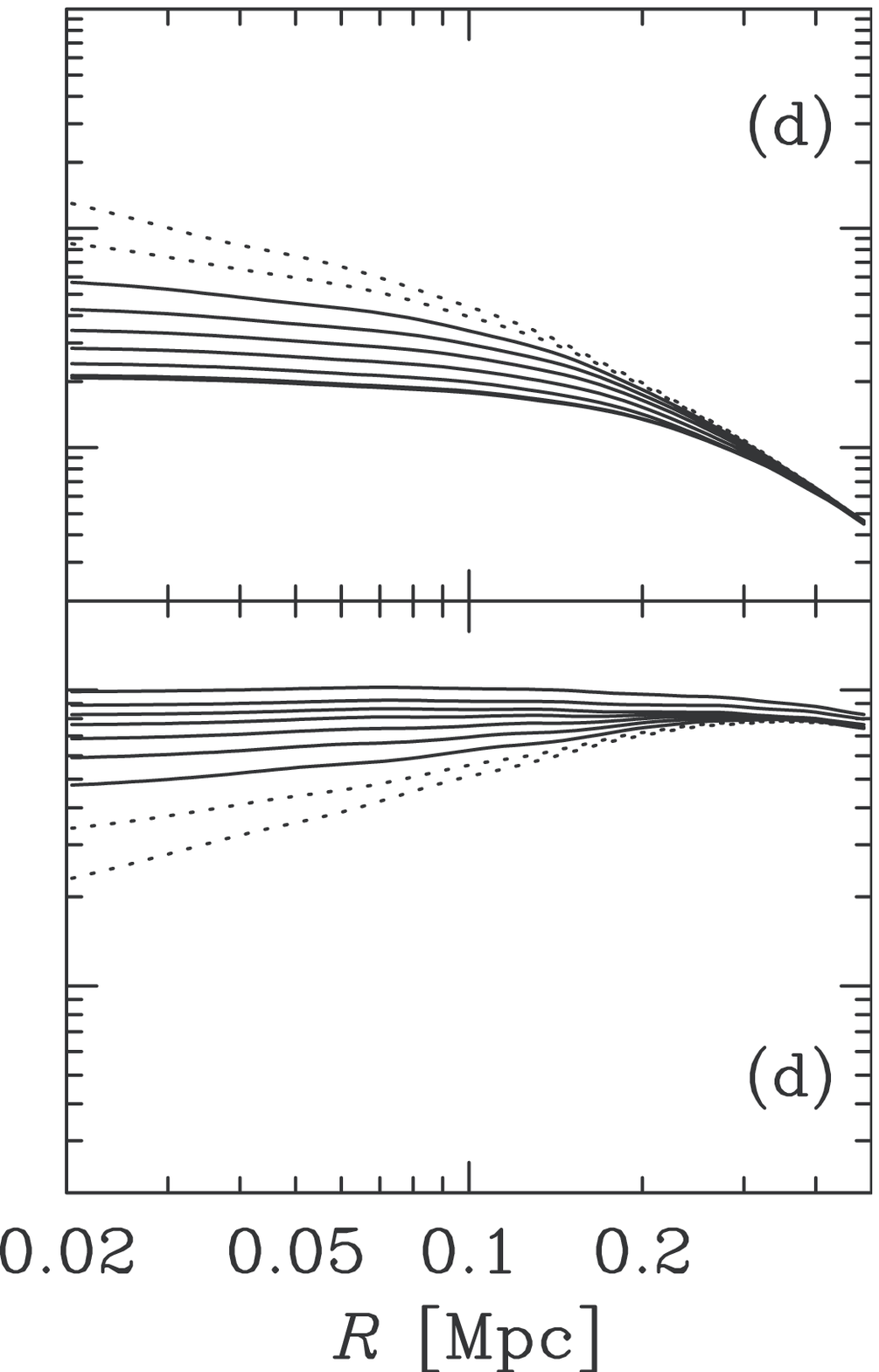}
\end{center}
\caption{Thermal evolutions of the density and temperature profiles of the four
simulated clusters. The lines are drew for intervals of about $\sim 200$~Myr
until $t_{\rm cool}$. Dotted lines represent the profiles for continuing
calculations over $t_{\rm cool}$, which are excluded for the analysis in figure 4.}
\label{fig:5}
\end{figure*}

\begin{figure}
\begin{center}
\FigureFile(77mm,54mm){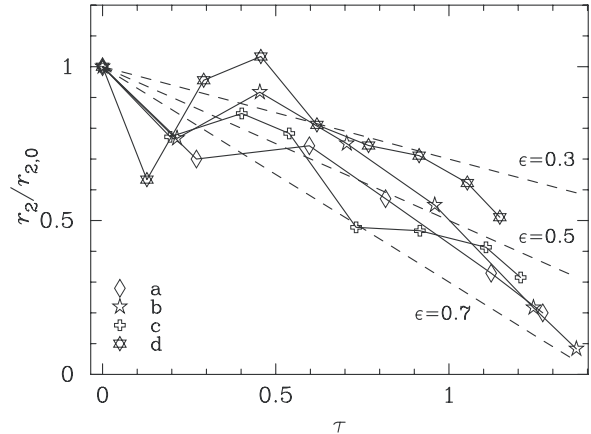}
\end{center}
\caption{Thermal evolutions of inner-core radii. The dashed lines represent
equation (\ref{eq:7}) with $\epsilon=0.3,~0.5,~0.7$. }
\label{fig:6}
\end{figure}

The evolutions of the density and temperature profiles (figure~\ref{fig:5})
imply the self-similar evolution in the core region, since the evolution is
predominant in the inner component.
We try to approximate the evolution
of inner-core radii with a parameter, $\epsilon$, and the normalized time,
$\tau\equiv t/t_{\rm cool}$, as
\begin{equation}\label{eq:7}
r_2=r_{2,0}(1-\epsilon\tau).
\end{equation}
Figure \ref{fig:6} shows that equation~(\ref{eq:7}) 
can represent the inner-core evolution, i.e. thermal
evolution of the small core can be described by the self-similar relation,
normalized by $t_{\rm cool}$ and $r_{2,0}$ which are obtained from the initial
outer component. This self-similarity may be seen also in the clusters having
different initial central densities, because in the $\beta$-model the central
density determines the amplitude of the profile while the flat-core region
clearly exists; the core radius is related mainly to the envelope slope (see
figure~\ref{fig:1}). 

%%% figure 6 here %%%

Such a self-similarity is, however, likely lost in the observed small cores. At
the initial state, the inner-core size can be represented by the outer-core
size, as
\begin{equation}\label{eq:8}
r_{2,0}=\alpha r_{1,0},
\end{equation}
where $\alpha\sim 0.3$ is obtained from figure~\ref{fig:3}.
While the thermal evolution is predominant in the inner component, outer core
is roughly constant, i.e. $r_1\simeq r_{1,0}$, so that $\delta r_2\equiv
r_{2,0}-r_2$ is written with equations~(\ref{eq:7}) and (\ref{eq:8}) as
\begin{equation}\label{eq:9}
\delta r_2\simeq\alpha\epsilon\frac{r_{1,0}}{t_{\rm cool}}t.
\end{equation}
From equations (10) and (11) in AM05, the initial central cooling time
is given by
\begin{equation}\label{eq:10}
t_{\rm cool}\propto\left(\beta\frac{x_{\rm vir}^2}{1+x_{\rm vir}^2}
\right)^{-1/2}r_{1,0}.
\end{equation}
Therefore, equation (\ref{eq:9}) is written as
\begin{equation}\label{eq:11}
\delta r_2\propto\alpha\epsilon\beta^{-1/2}t
\end{equation}
for the clusters with $x_{\rm vir}\gg 1$ (AM05).
Equation (\ref{eq:11}) suggests that the
self-similar relation in the small cores may be lost for not only the different
age of the cluster from the last major merger, i.e. the time when the cluster
formed to the current size and down to the hydrostatic equilibrium, which is
labeled by $t$, but the various values of $\beta$ from $\sim 0.6$ to $\sim 1.0$
in observed clusters as mentioned in AM05. 

\subsection{Evolution of Hydrostatic Structure}

\begin{figure}
\begin{center}
\FigureFile(77mm,54mm){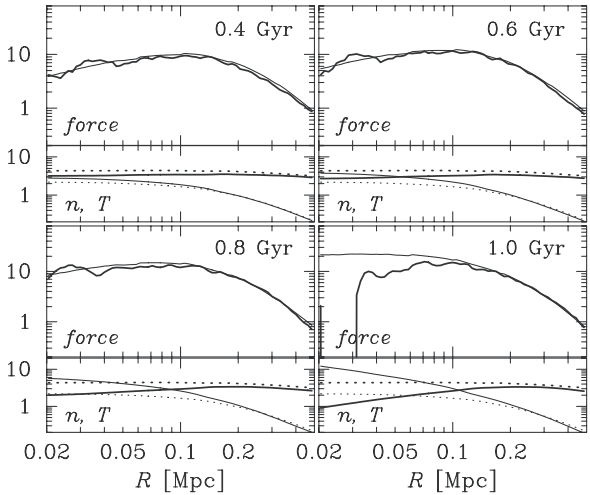}
\end{center}
\caption{Hydrostatic balance of cluster (b) after 0.4~($0.5~t_{\rm cool}$),
0.6~($0.75~t_{\rm cool}$), 0.8~($t_{\rm cool}$), and 1.0~($1.25~t_{\rm
cool}$)~Gyr. The lines are the same descriptions as figure~\ref{fig:2}. The
dotted lines are the initial values of the density and temperature. }
\label{fig:7}
\end{figure}

\begin{figure}
\begin{center}
\FigureFile(77mm,54mm){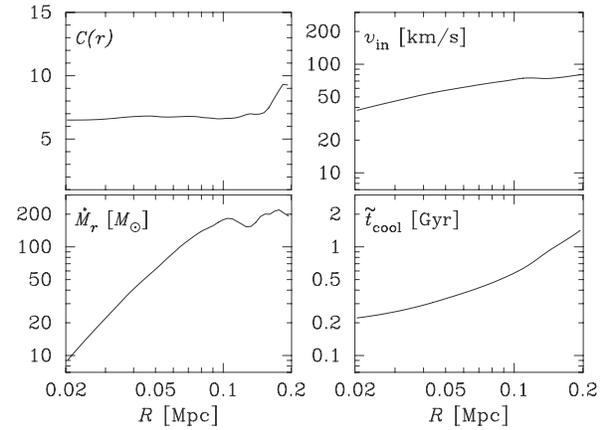}
\end{center}
\caption{$C(r)$~$(\equiv(\tilde{M}_r/\tilde{t}_{\rm cool})/\dot{M}_r$), the
infall velocity ($v_{\rm in}$), the mass flow rate ($\dot{M}_{r}$), and the
cooling time ($\tilde{t}_{\rm cool}$) after 0.8~Gyr ($t_{\rm cool}$) of cluster
(b).  }
\label{fig:8}
\end{figure}

Calculations show that the cooling gas approximately keeps the hydrostatic balance
between the gravitational force and pressure gradient at each step until $\sim
t_{\rm cool}$ (figure~\ref{fig:7}).  We compare our result with
quasi-hydrostatic cooling discussed by Masai and Kitayama (2004).  They suggest
that $\dot{M}_r\sim \tilde{M}_r/\tilde{t}_{\rm cool}$ at every $r$ or 
\begin{equation}\label{eq:12} C(r)\equiv(\tilde{M}_r/\tilde{t}_{\rm
cool})/\dot{M}_r\sim {\rm constant},
\end{equation} where $\dot{M_r}=4\pi \rho_{\rm g} r^2dr/dt$ is the continuity
equation,
$\tilde{M}_r=4\pi r^3\rho_{\rm g}/3$ and 
$\tilde{t}_{\rm cool}=3n_{\rm g}kT/({n_{\rm g}}^2\Lambda)$ are the mass of a
uniform gas sphere and the cooling time for the local values at radius $r$,
respectively.  As shown in figure~\ref{fig:8}, our calculation shows $C(r)$ is
nearly constant. Until $\sim t_{\rm cool}$, constant $C$ is seen commonly for
all the model clusters, not only the King case but also the NFW case. 

%%% figure 7 here %%%

Masai and Kitayama (2004) show some properties of quasi-hydrostatic cooling by
approximating the density/temperature profiles in the power-law form.  We
examine our results in the same manner with $\rho_{\rm g}\propto r^{\alpha}$,
$T\propto r^{\eta}$ and inflow velocity $v_{\rm in}\propto r^\zeta$. Since
$C(r)\propto r^{1+\alpha-\eta/2-\zeta}$, quasi-hydrostatic cooling means
$1+\alpha-\eta/2-\zeta\sim 0$ or
\begin{equation}\label{eq:13}
\alpha\sim -1+\eta/2+\zeta.
\end{equation}
Such a relation is actually found at $\sim 40$~kpc:
$\alpha=-0.42$ is comparable to $-1+\eta/2+\zeta\sim -0.41$ with ($\eta$,
$\zeta$)=(0.28, 0.45). At $\sim 70$~kpc, however, $\alpha=-0.63$ is smaller
than $-1+\eta/2+\zeta\sim -0.54$ with ($\eta$, $\zeta$)=(0.25, 0.34).
Therefore, quasi-hydrostatic balance may be satisfied marginally.  When the
cluster center cools rapidly at $t>t_{\rm cool}$ (figure~\ref{fig:5}), the
balance breaks because a large amount of inflow is required to maintain the balance.
This also leads to the fact that the spatial resolution in calculations becomes
practically worse than estimated simply from the number of SPH particles.

%%% figure 8 here %%%

At $r\lesssim 100$~kpc, the gas inflow velocity $v_{\rm in} \sim 0.1 c_s$,
where $c_s$ is the sound speed. The mass inflow rate $\dot{M}_r({\rm
100~kpc})\sim 200~\MO$ is decreasing to $\dot{M}_r({\rm 20~kpc})\sim 10~\MO$,
which are much smaller than expected for the classical cooling flow model and
consistent with the quasi-hydrostatic cooling model by Masai and Kitayama
(2004). It should be noted that they consider $\zeta < 0$ for smooth inflow,
but we find in figure~\ref{fig:8} that the inflow velocity gradually decreases toward the
cluster center, i.e. $\zeta>0$ rather than $\zeta<0$. It is, however,
consistent with their prediction that $\tilde{t}_{\rm cool}\propto r^{1-\zeta}$
is well satisfied in $r<100$~kpc.  

$\zeta>0$ implies that the gas is heated by viscosity.
We confirmed the convergence of the solution by using SPH-particles 
greater than $\sim 30,000$, which is also sufficient resolution to remove
improperly accelerating cooling (\cite{SH02}).
The gas inflow might be exaggerated by underestimating the viscous heating
at the dead center of the cluster, though the influence reaches
up likely to $\sim 2h$ from the center because SPH-particles within $2h$ at a
given point contribute to the smoothed estimate in the present work.
Therefore, at each step until $\sim t_{\rm cool}$, quasi-hydrostatic balance
is likely attained without some heat sources such as electron conduction
and/or AGN activities. 

\section{Concluding Remarks}

We investigate the thermal evolution of cluster cores: the core size of the
intracluster gas can vary as the gas cools radiatively and flows toward the
cluster center. In order to compare with the observed core radii, we apply the
$\beta$-model or double $\beta$-model, as done by the observational studies, to
analyze the gas profile, although the model is not always good for cooling
cores of more or less center-peaked profiles.  Thermal evolution of the gas may
be classified into the following three stages: (I) at $\tau\equiv t/t_{\rm
cool}\lesssim 0.5$, the profile is represented by the single $\beta$-model with
a large core or by the outer-core dominated double $\beta$-model, (II) at $0.5
\lesssim \tau \lesssim 1$, the profile is represented well by the double
$\beta$-model, and (III) at $\tau \gtrsim 1$, the profile is represented by the
single-$\beta$ model with a small core or by the inner-core dominated double
$\beta$-model.

Until $\sim t_{\rm cool}$, which is the central cooling time for
the initial gas profile, the gas cools keeping the hydrostatic balance between
the gravity on the gas and the gradient of the thermal pressure. This evolution
is explained by the quasi-hydrostatic cooling model proposed by Masai and
Kitayama (2004). The properties of the cooling gas in our calculations, such as
the mass inflow rate, radius dependence of the local cooling time and constant
$C$, are consistent with this model in the regime where quasi-hydrostatic
structure is attained. We find that the inflow velocity of the gas decreases
toward the cluster center, and its mass flow rate is about $10~\MO$ at 20~kpc.
Quasi-hydrostatic condition is marginally satisfied up to $t \sim t_{\rm cool}$,
and then the cluster center cools rapidly. 
Regarding the relation with the core-size distribution, it
may be an important clue for understanding thermal properties of clusters that
even the small core peak is produced by clusters of $t \lesssim t_{\rm cool}$,
in other words, by clusters that are still keeping quasi-hydrostatic balance. 

Analyzing the simulated clusters
with the single and double $\beta$-models, we demonstrate the core size
distribution exhibits two distinct peaks at $\sim 100$--300~kpc and at $\sim
30$--100~kpc with a valley at $\sim 120$~kpc until $t_{\rm cool}$ and is in
fairly good agreement with the observed distribution (Ota and Mitsuda 2002;
Akahori and Masai 2005).  This implies that the origin of small cores can be
explained by radiative cooling or thermal evolution of the gas.  However, a
question remains: the self-similarity is kept in simulated clusters through the
thermal evolution with cooling, but seems to be lost in the observed small core
clusters for the various values of $\beta$ from $\sim 0.6$ to $\sim 1.0$
(Akahori and Masai 2005). 

\bigskip

The authors would like to thank Tetsu Kitayama for his help for computation
resources, and him and Naomi Ota for their useful discussions.


\begin{thebibliography}{99}
\bibitem[Akahori, Masai(2005)]{AM05}
	Akahori,~T., \& Masai,~K. 2005, \pasj, 57, 419
\bibitem[Cavaliere, Fusco-Femiano(1976)]{CF76}
		Cavaliere,~A., \& Fusco-Femiano,~R. 1976, \aap, 49, 137
\bibitem[Ettori et al.(2004)]{Ettori04}
		Ettori,~S., Tozzi,~P., Borgani,~S., \& Rosati,~P. 2004, \aap, 417, 13
\bibitem[Fabian(1994)]{Fabian94}
		Fabian,~A.~C. 1994, ARA\&A, 32, 277
\bibitem[Hernquist, Katz(1989)]{HK89}
		Hernquist,~L., \& Katz,~N. 1989, \apjs, 70, 419
\bibitem[King(1966)]{King66}
		King,~I.~R. 1966, \aj, 71, 64
\bibitem[Lewis et al.(2002)]{Lewis02}
		Lewis,~A.~D., Stocke,~J.~T., \& Buote,~D.~A. 2002, \apj, 573, L13
\bibitem[Makishima et al.(2001)]{Makishima01}
		Makishima,~K., et al. 2001, \pasj, 53, 401
\bibitem[Masai, Kitayama(2004)]{MK04}
		Masai,~K., \& Kitayama,~T. 2004, \aap, 421, 815
\bibitem[Mohr et al.(1999)]{Mohr99}
		Mohr,~J.~J., Mathiesen,~B., \& Evrard,~A.~E. 1999, \apj, 517, 627
\bibitem[Monaghan(1992)]{Monaghan92}
		Monaghan,~J.~J. 1992, ARA\&A, 30, 543
\bibitem[Monaghan, Gingold(1983)]{MG83}
		Monaghan,~J.~J., \& Gingold,~R.~A. 1983, J. Comput. Phys., 52, 374
\bibitem[Navarro, Frenk, and White(1996)]{NFW96}
		Navarro, J. F., Frenk, C. S., \& White, S. D. M.,1996 \apj, 462, 563
\bibitem[Ota, Mitsuda(2002)]{OM02}
		Ota,~N., \& Mitsuda,~K. 2002, \apj, 567, L23
\bibitem[Ota, Mitsuda(2004)]{OM04}
		Ota,~N., \& Mitsuda,~K. 2004, A\&A, 428, 757
\bibitem[Peterson et al.(2001)]{Peter01}
		Peterson,~J.~R., et al. 2001, \aap, 365, L104
\bibitem[Ruszkowski, Begelman(2002)]{RB02}
		Ruszkowski,~M., \& Begelman,~M.~C. 2002, \apj, 581, 223
\bibitem[Sarazin(1986)]{Sarazin86}
		Sarazin,~C.~L. 1986, Rev. Mod. Phys. 58, 1
\bibitem[Springel, Hernquist(2002)]{SH02}
		Springel, V., \& Hernquist,~L. 2002, \mnras, 333, 649
\bibitem[Sutherland, Dopita(1993)]{SD93}
		Sutherland,~R.~S., \& Dopita,~M.~A. 1993, \apjs, 88, 253
\end{thebibliography}
\end{document}